# Secure User Authentication & Graphical Password using Cued Click-Points

Miss.Saraswati B.Sahu [#1], Associate Prof. Angad Singh [*2]

[#]*M.Tech Scholar & Information Technology & RGPV*
*NIIST ,Bhopal,India*

*Abstract*: The major problem of user registration, mostly text base password, is well known. In the login user be inclined to select simple passwords which is frequently in mind that are straightforward for attackers to guess, difficult machine created password mostly complicated to user take in mind. User authenticate password using cued click points and Persuasive Cued Click Points graphical password scheme which includes usability and security evaluations.
   This paper includes the persuasion to secure user authentication & graphical password using cued click-points so that users select more random or more difficult to guess the passwords. In click-based graphical passwords, image or video frame that provide database to load the image, and then store all information into database. Mainly passwords are composed of strings which have letters as well as digits. Example is alpha-numeric type letters and digits.

## 1. INTRODUCTION

There are a lot of effects that are most well-known about passwords; such as that user cannot memorize complicated password which is simple to identify [1-6].

A user registration is supposed to support strong passwords for preserve to keep in mind and protection. The user registration process allows choosing while influencing users proposed for difficult passwords. The task of selecting weak passwords is more monotonous, avoids users from making like choice. This type of registration schemes it is use for very protected password. Instead of increasing the burden on users, it is much easier to use the system's suggestions for a secure password - a feature absent in most schemes.

In this approach to create the primary normal click point visible password and cued click point created an inter lab connecting study with a variety of participants. The results shows and explained that Cued Click Points method is useful at condensed the number of user connected areas of the image where users are more likely to select click points while still maintaining usability. In this paper also analyze the efficiency of tolerance value and security rate. While we are not at variance that graphical passwords are the best approach to authentication, in research finding that it provide an excellent environment for exploring strategies for helping users select better passwords since it is easy to compare user choices. This method completely use simple of the user about his selected images. If user cannot define the total area of the image, the user can't register even though he is a authentic user. The user feels comfortable with this type of graphical passwords when compared with text passwords.

## 2. BACKGROUND

Instead of text password, Graphical passwords have been proposed to improve both usability and security issues. For registration process the text based password is better, while it is not more secure. By the use of bio metric or token base password is weak and problem oriented [7-9].In this method to moderate the troubles with conventional methods. Greg Blonder first described the idea of graphical password in the year 1996. In the use of Graphical passwords that provides alternative, and are the center of this paper.

### 2.1 Graphical Passwords Using Normal Click Points

The password which is secure information for every user if the password is used like graphical is use for user registration process. Graphical password is attempts to influence the user stored in mind for visible data. Previous survey of visible passwords is obtainable in a different place of interest in this are normal click click-based graphical passwords. In such systems, users identify and target formerly preferred locations within more than one images. Example systems include Pass Points [15] and Cued Click- Points (CCP) [10].

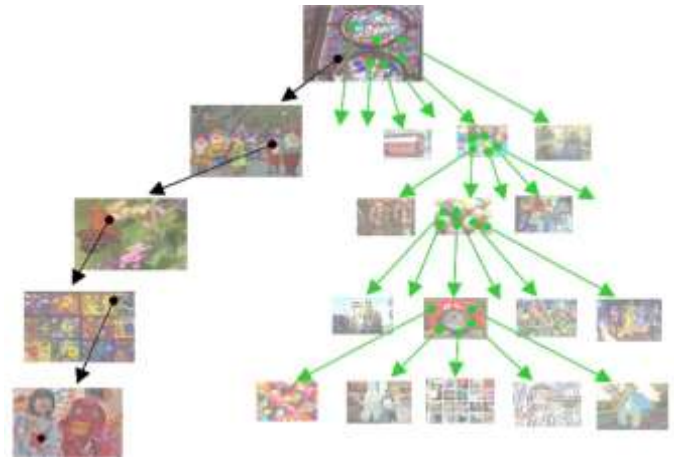

Fig.1. click point password for user navigation process. The next image determine with each click. [11]

The Click Points is a planned to set pass point. The method of click-point user selects single points instead of 3 point on 3 images. The method provides cued-recall the facilities visible cued immediately provide message if they enter wrong click point.

Using the previous survey Persuasive technology is to inspire to encourage new concept. This method provides solution to user for selecting best password. User will not avoid the effective elements as well as final password must be memorable. The path for candidate to choose difficult password which is can't to access easily. If the click point select as randomly so formation of hotspot is minimized.

In Pass-Points, the user password use continues five click points provides on particular image. A User might choose the appropriate pixel for password. Mostly pass point is frequently used [15-16], if the given security is poor then password is quite simple for attackers to crack. In the method of various define click point [17] that make for simplify pattern for helpfulness for hotspot for attackers. Instead of five click point on single image the cued click point method apply only single click point on various five images, which presented as continues images. The display of next image is depending on previously image. In the fig 1 there are path provides for all image set. In this user selected given image which is amount of their click point provided.

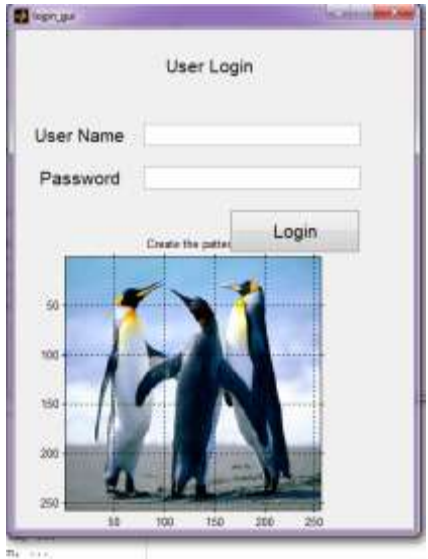

Fig. 2 Create the Pattern on image and login

### 3 PERSUASIVE CUED CLICK POINTS

In previous approach describe the weak password is mostly possible only in case of hotspot and lower security. If attackers is to determine and prioritize higher probability passwords for more appropriate guessing attacks. Basically survey shows those various user highlighted the common area on that image. In this case user choose their own click-based graphical passwords without direction, hotspots will remain an issue. Final thing is that to satisfies strong security to providing reduced number of choices it extra overriding and very much difficult. In result, the safe selection root of least resistance defined by user [2].the selection of this method [7], proposed the user make a password, User select image from database and choose pattern as password and they choose a click point that to selected area for click point, an area on image button appears only during password creation. Normally next entry of password the image generated simply without the shading of that view point so user can click anywhere on the images. In the click point and pass-point the login define the tolerance squares of main points.

I. In the first step user feasible to pick click-points that put fervent to recognized hotspots.
II. Click-point share diagonally users will be additional randomly separate and will not form latest hotspots.
III. User login security success rates will be better to those of the original CCP system.
IV. Login security success rates will boost, when receipt value is lower value.
V. User will get their passwords which are more protected with PCCP than participants of the original CCP system.

Password space for a system is the all number of unique password will generate depend on system rules. Normally, complicated of long password is more difficult to identify to given actual password. In the method of click point password space is (w x h )/t$^2$)c in this the size is in pixel (w x h) is separated with the size of tolerance square (t$^2$), the total number of tolerance squares/ image, raised to power of number of click point provided in image password.

### 4. SYSTEM DESIGN

The system designed with the help of three modules such as Authentication Process, User registration /Sign Up process modules , Loading video and creating frame module, picture selection module and system login module (see Figure 3).

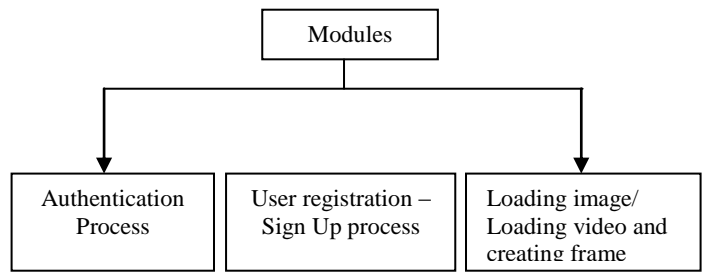

Fig. 3.System design modules

User authentication process user select login screen ask user id and password, then user will retrieve the image from data base and ask user for click point here minimum 3 point are necessary require, this process is continuously repeat until the no match found , so user will continue to re- login into the page.

In user registration module user enter the user name in user name field and also suitable tolerance value. When user entered the all user details in registration phase, these registration data of different users stored in data base and used during login phase for verification. In the previous section select image from system and create minimum 3 click points then store the total information of user to the data base and ask the user to select more images. Otherwise ask for load the video and select a particular frame from the video and provide object name, finally store the information into data base.

Two Types of flowchart:
  I.  User Authentication Process/ Login flow chart
  II. User registration-Sign up Process flow chart

Below flowchart (see Figure 4 & 5) defines the user registration define main procedure, this procedure include two registration phase (user ID) and picture selection phase. The flow of the given process generates from registering user id and clearance value. Once user completes all the user details then precede to next stage, which are given selecting click points on generated images, which ranges from 1-5. After done with all these above procedure, if clue is define then frame will be created.

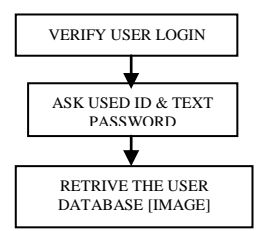

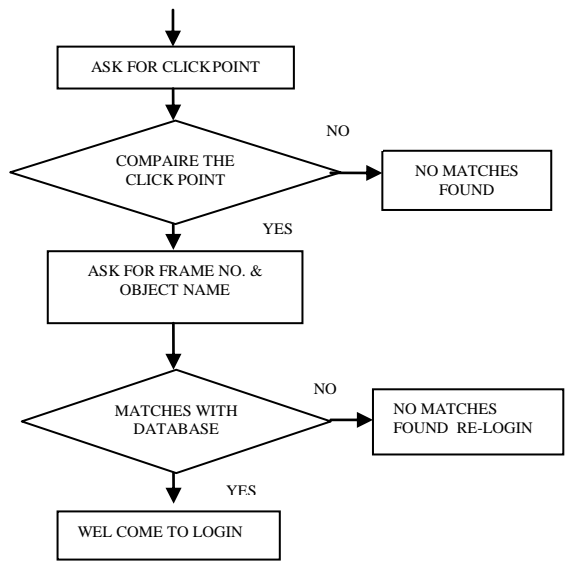

Fig.4. Authentication Process/ Login

In picture selection phase user select any image as passwords and provides continues of three click points on image. Any pixel on the image can choose for the password by user. The user select image from database. Then asking for click point then asking for frame number of object, if the appropriate thing is match with the database then user show the login screen otherwise the user must be re enter the user id and password.

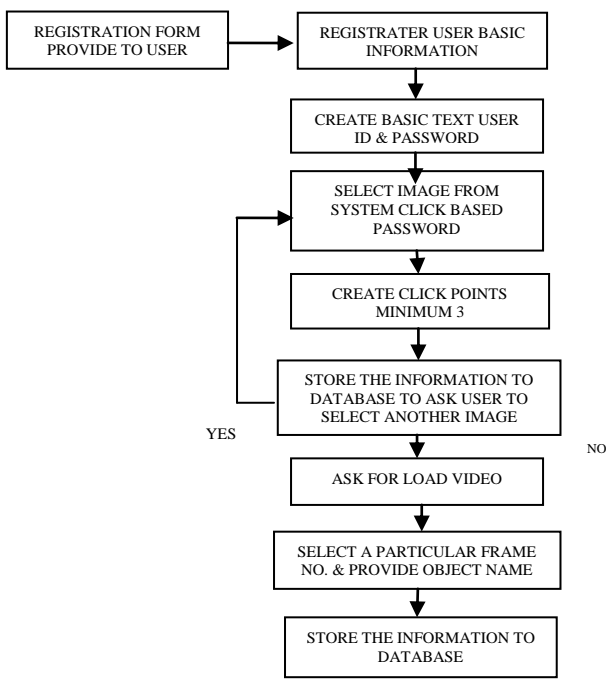

Fig.5. User registration –Sign Up process

After selecting procedure, the click point normal click base or Persuasive Cued Click-Points, next step to upload image or video. Here image can be taken from any location and video is predefined, and then create a single frame. During system login, the user must repeat the procedure of login otherwise they will be getting no match found message.

## 5. IMPLEMENTATION

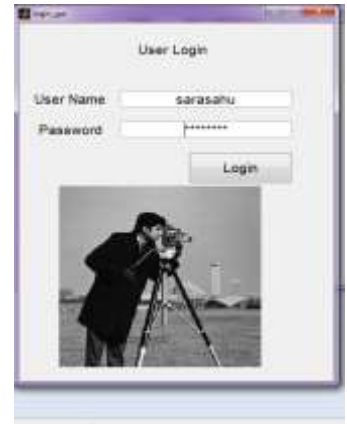

Fig 6 main menu user registration process login, signup, forgot password

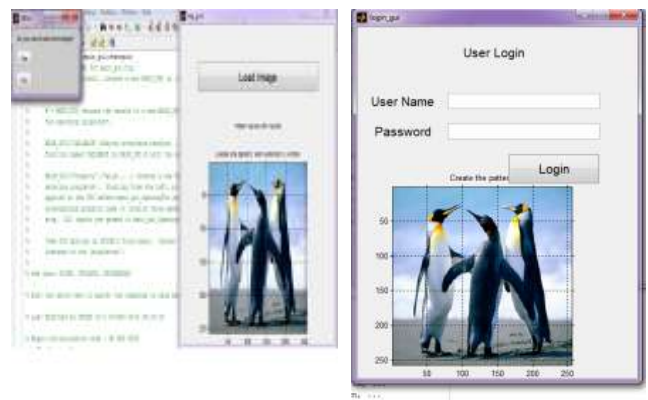

Fig.7 Creating Password interface using Cued click points

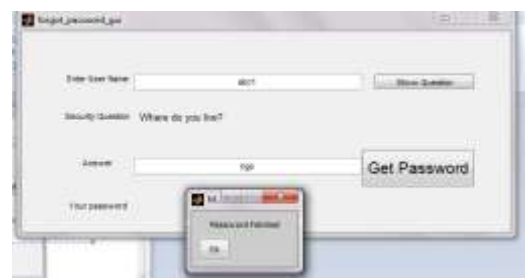

Fig. 8 security question if user forgot password

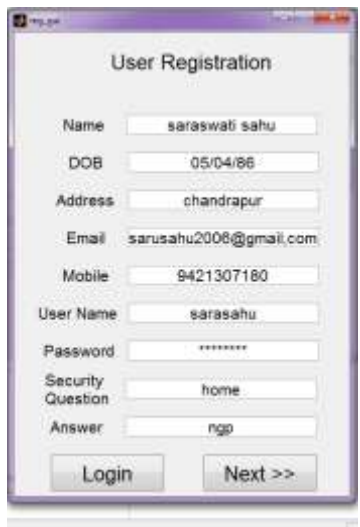

Fig.9 sign up process if user is not registered

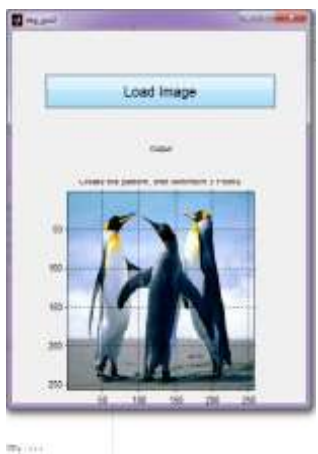

Fig.10 loading more images for cued click point connects

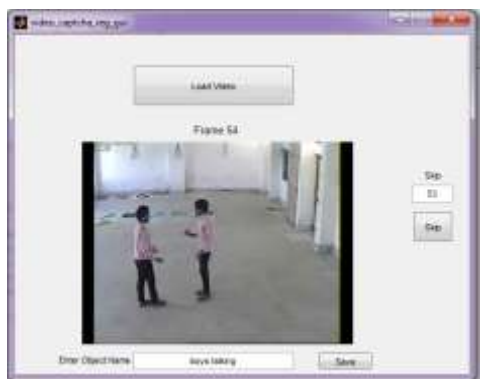

Fig.11 loading video and creating frame

## 6. RESULTS AND ANALYSIS

The study and analysis of normal point( CCP) and persuasive cued click point (PCCP) was designed to explore ways of increasing the efficiency of tolerance value and also conducted lab study for comparison between login success rate and security success rate of existing CCP's and proposed PCCP's.

*A. Efficiency of the tolerance value*

Initially eight participants are considered for the experiment. Each participant has a unique password which includes clicking on 5 click points in 5 different images. Each image contain of different characters (image details), among which the participant needs to click on any one point of his choice to make it a click point in the series. Similarly the participant select a click point each of the images. Then, the participant logs in with that password, meantime the other participants are made to stand in a group behind the participant who is entering the password and are made to peek in over the shoulder of the participant and observe his password (the click points on the images). The first participant has logged out once, the other participants are asked to enter the same password which they have observed of the first participant.

**Tolerance value**: It is the value which defines the degree of closeness to the actual click point.

**Tolerance region**: The area describes an original click point is accepted as correct since it is unrealistic to expect user to accurately target an exact pixel.

**Success rate**: It is the rate which gives the number of successful trails for a certain number of trials, the success rates calculated as the number of trails given completed without errors or restarts.

**Shoulder surfing**: It is the process by which the person standing behind the person entering the password observes the password. It is a type of capture attack. This attack occurs when attackers are directly obtaining the passwords by intercept the user put values and password for that.

The below table 1 shows the result of the tolerance value efficiency of the PCCP method.

| No | Tolerance value | Success rate | Percentage of success rate | Security (in percentage) |
|---|---|---|---|---|
| 1 | 5 | 7/8 | 87.5 | 12.5 |
| 2 | 4 | 6/8 | 75.8 | 37.5 |
| 3 | 3 | 3/8 | 37.8 | 62.5 |
| 4 | 2 | 2/8 | 25 | 75 |
| 5 | 1 | 0/8 | 0 | 100 |

## 7. POSSIBLE SECURITY ATTACK

Here the number of possible attack regarding to security concern.

### 7.1 Password guessing attack

Most critical guessing attack against Cued Click Points is a brute force attack, through the predictable achievement behind explore partially of the password space yet, skewed password separation that provides attacker to give on this harm for that metadata of information can be formed of server-side information. Then this dictionary information can be used for the guessing of the click point in an image.

### 7.2 Password Capture attacks

In this attacks when attackers directly obtain passwords by intercepting user entered data and tricking users into revealing their passwords. In capturing one sign in instance allows deceptive access by a simple replay attack. Some security schemes are vulnerable to carry surfing threat. Examine the estimated location of click points may reduce the figure of guesses essential to decide the user's password. User crossing point, normally attacks against cued-recall graphical passwords, a frame of orientation must be recognized between parties to correspond the password in enough detail. Graphical passwords may also be shared by taking picture, capturing screen shots, or drawing, although requiring more attempt than for text passwords.

### 7.3 Image Pattern Attack

In this pattern-based attack that provides passwords with of click-points ordered in a consistent horizontal and vertical direction (with straight lines in any direction, arcs, and step patterns), as well as ignores any image-specific characteristics such as hotspots. Given that Cued Click Point passwords are fundamentally impossible to tell apart from arbitrary for click-point distributions among the x- and y-axes, angles, shapes and slopes.

### 8. CONCLUSION

As we are providing the only one image for the authentication purpose it is easier for the user to remember and difficult for the attacker to attack because it is difficult for the attacker to see at click points area of the image. Using of the graphical password method it is undeveloped. In the various research and study the various approach the graphical password method is provide wide strong security and very convenience.

By using database image or creating a frame from video password become strong. This approach encourages and guides users in selecting more random click-based graphical passwords. A key feature in this method is that creating a secure password is the given path-of-least-resistance, it likely to be more productive than schemes where it securely adds an extra load on users.